\documentclass[sigconf]{acmart}
\usepackage{booktabs} 
\usepackage[ruled,vlined]{algorithm2e}
\usepackage[noend]{algorithmic}
\usepackage[inline]{enumitem}
\usepackage{tabulary}
\usepackage{tabularx}
\usepackage{diagbox}
\usepackage{makecell}
\newcolumntype{x}[1]{>{\centering\arraybackslash}p{#1}}
\usepackage{array}
\usepackage{multirow}
\usepackage{float}
\usepackage{epstopdf}
\usepackage{amssymb}
\usepackage{amsmath,amsthm,amssymb,mathtools}
\usepackage{xcolor}
\usepackage{graphicx}
\usepackage{comment}
\SetKwProg{Fn}{Function}{}{}
\usepackage{tablefootnote}
\usepackage{tabularx}
\usepackage{threeparttable}

\renewcommand{\algorithmicrequire}{\textbf{Input:}}

\newenvironment{myitemize}{\begin{list}{$\bullet$}
{\setlength{\topsep}{1mm}
\setlength{\itemsep}{0.25mm}
\setlength{\parsep}{0.25mm}
\setlength{\itemindent}{0mm}
\setlength{\partopsep}{0mm}
\setlength{\labelwidth}{15mm}
\setlength{\leftmargin}{4mm}}}{\end{list}}

\DeclarePairedDelimiter{\ceil}{\lceil}{\rceil}
\def\BibTeX{{\rm B\kern-.05em{\sc i\kern-.025em b}\kern-.08em
    T\kern-.1667em\lower.7ex\hbox{E}\kern-.125emX}}

\begin{document}

\title{Leveraging Weakly-hard Constraints for Improving System Fault Tolerance with Functional and Timing Guarantees}





\author[]{Hengyi Liang, Zhilu Wang, Ruochen Jiao, Qi Zhu}
\email{{hengyiliang2018@u.,zhilu.wang@u.,RuochenJiao2024@u.,qzhu@}northwestern.edu}
\affiliation{
\institution{Northwestern University}
\city{Evanston}
\state{Illinois}}


\begin{abstract}
Many safety-critical real-time systems operate under harsh environment and are subject to soft errors caused by transient or intermittent faults. It is critical and yet often very challenging to apply fault tolerance techniques in these systems, due to their resource limitations and stringent constraints on timing and functionality. 
In this work, we leverage the concept of weakly-hard constraints, which allows task deadline misses in a bounded manner, to improve system's capability to accommodate fault tolerance techniques while ensuring timing and functional correctness. 
In particular, we
\begin{enumerate*}[label={\alph*)},font={\bfseries}]
    \item quantitatively measure control cost under different deadline hit/miss scenarios and identify weak-hard constraints that guarantee control stability;
    \item employ typical worst-case analysis (TWCA) to bound the number of deadline misses and approximate system control cost;
    \item develop an event-based simulation method to check the task execution pattern and evaluate system control cost for any given solution; and 
    \item develop a meta-heuristic algorithm that consists of heuristic methods and a simulated annealing procedure to explore the design space.
\end{enumerate*}
Our experiments on an industrial case study and a set of synthetic examples demonstrate the effectiveness of our approach.
\end{abstract}

\maketitle
\fancyhead[]{}

\section{Introduction}
Many real-time embedded systems, such as automotive, avionics, and industrial automation systems, often operate under harsh environment and are subject to soft errors caused by transient or intermittent faults (e.g., those from radiation~\cite{baumann2005radiation}). As those systems are often safety-critical, it is important to improve their resiliency by applying fault tolerance techniques~\cite{anderson2983framework,kumar2011fault}.




In the literature, various error detection and recovery mechanisms have been proposed~\cite{zheng2015analysis,gao2013using,izosimov2009analysis,pop2009design}. For instance, to address soft errors, there are both hardware based approaches~\cite{rehman2018hardware,Bastos2020,rajabzadeh2005hardware} and software approaches~\cite{gao2013using,miremadi1992two,oh2002control}. In this work, we focus on addressing transient soft errors through software layer, by relying on error detection techniques to detect potential soft errors and possibly performing recovery jobs to correct them. 
As defined in~\cite{gao2013using}, there are two main categories of error detection techniques, i.e., embedded error detection (EED) and explicit output comparison (EOC). EED-type techniques have built-in error detection mechanisms and do not reply on redundant execution. Some common EED approaches include watchdog timer~\cite{miremadi1992two}, control flow checking and instruction signature checking~\cite{oh2002control}. In contrast, EOC-type techniques rely on explicit redundant execution with either temporal redundancy or spatial redundancy, e.g., executing the same task at least twice and compare the outputs. 
One common approach of EOC is the triple modular redundancy scheme~\cite{lyons1962the}. In this work, we consider the general type of EED techniques and an EOC technique based on temporal redundancy, i.e., EOC tasks are executed twice on the same computation resource and in the case of a soft error, a re-execution job is scheduled immediately on the same resource.

Both EED and EOC techniques incur significant timing overhead, 
and thus quantitative schedulability analysis is needed to ensure system timing correctness. For instance, the work in~\cite{gao2013using} presents an offline scheduling algorithm for EOC-type techniques. The work in~\cite{zheng2015analysis} explores the tradeoff between EOC- and EED-type techniques and presents an algorithm to optimize their selection and scheduling, while considering timing constraints. However, 
applying fault tolerance techniques to resource-constrained real-time systems is quite  challenging and sometimes infeasible, as it is often difficult to meet the stringent hard timing constraints with the additional overhead from those fault tolerance techniques.

In this work, we present a novel approach for improving system fault tolerance that leverages the concept of \emph{weakly-hard constraints} as defined in~\cite{Bernat_TC_01}, where bounded deadline misses are allowed, to provide more slack in task execution and enable the addition of more error detection and correction measurements. 
Unlike traditional hard real-time constraints, weakly-hard constraints allow occasional deadline misses in a bounded manner, which are often specified as the maximum number of deadline misses allowed within a given number of consecutive job instances (or a window of time)~\cite{Bernat_TC_01, hamdaoui1995dynamic}. 

The exploration of weakly-hard constraints is motivated by the fact that many system functions (e.g., control or sensing functions) can tolerate certain degree of deadline misses while still satisfy functional correctness requirements. For example, recent works have studied control performance and stability under deadline misses specified by weakly-hard constraints~\cite{pazzaglia2018beyond,Goswami_TCST14,huang2019formal}. In~\cite{Goswami_TCST14}, the authors prove an analytical upper bound of deadline miss ratio to ensure the stability of a distributed embedded control platform. In~\cite{pazzaglia2018beyond}, the authors study the impact of deadline miss pattern on control performance. In~\cite{huang2019formal}, the authors present a method to formally verify the safety of certain control systems under weakly-hard constraints. 
On the other hand, a number of approaches have been presented for schedulability analysis of real-time system with weakly-hard constraints~\cite{Bernat_TC_01, Sun_TECS_17, Quinton_DATE_12, Xu_ECRTS_15, li2006providing}. In~\cite{Bernat_TC_01}, the response time analysis for periodic task is discussed. In~\cite{Sun_TECS_17}, the authors model the schedulability analysis as a mixed integer linear programming (MILP) problem and apply it to periodical tasks with unknown task activation offset. In~\cite{Quinton_DATE_12}, a model is proposed to describe task activation pattern, and typical worst-case analysis (TWCA) is introduced to bound the number of deadline misses due to overload. The work in~\cite{Xu_ECRTS_15} further improves the approach from~\cite{Quinton_DATE_12}.
Then, there is also limited work on trying to leverage the scheduling flexibility from weakly-hard constraints to improve other design objectives. For instance in~\cite{liang2019security}, a co-design approach is presented to improve system security while ensuring control safety. 

Our work is the \textbf{first to leverage weakly-hard constraints for improving fault tolerance}. There are two unique challenges to address for solving this problem: 1) While exploring weakly-hard constraints, we have to ensure that the allowed deadline misses will not cause functional incorrectness. In this work, we focus on the stability of control tasks under deadline misses, and the behavior of these tasks is particularly difficult to analyze when we consider the possible faults on them. 2) We need to analyze the system schedulability under the possible deadline misses from weakly-hard constraints \emph{and} the potential redundant task execution from fault tolerance techniques. Addressing these two challenges requires new methods for both control and schedulability analysis. 

We address these two challenges by developing new methods to analyze control stability and system schedulability under deadline misses, faults, and the application of EED or EOC fault-tolerance techniques. Based on these analysis methods, we also develop an optimization algorithm for exploring the design space to improve a system-level fault-tolerance metric. More specifically, our work makes the following novel contributions:  
\begin{myitemize}
    \item We develop a control analysis method for  linear time-invariant (LTI) systems to formally derive the weakly-hard constraints that can ensure system stability (e.g., the system can be brought back to the equilibrium state under deadline misses), and for quantitatively measure the control cost under different deadline hit/miss patterns. 
    \item We develop two schedulability analysis methods. One is to model tasks as the superposition of typical and overload activation and provide an upper-bound of the deadline misses (the control cost can be approximated based on this upper-bound). The other method uses an event-based simulation to record the exact pattern of deadline hits and misses (the worst-case control cost can be calculated under single transient error in this method).
    \item  We develop a meta-heuristic optimization algorithm to explore the design space, including task allocation, priority assignment, and the choice of fault tolerance techniques (EED, EOC, or none). We conduct experiments on an industrial case study and a set of synthetic examples. Our experiments demonstrate the effectiveness of our approach in improving system fault tolerance and trading off between control cost and error coverage.
\end{myitemize}

The rest of the paper is organized as follows. Section~\ref{sec:System_overview} introduces our system mode, including task execution model and control model. Section~\ref{sec:prob_form} presents our problem analysis and formulation, including the analysis on control stability and cost. Section~\ref{sec:meta_heuristic} introduces our schedulability analysis methods and our meta-heuristic optimization algorithm.
Section~\ref{sec:experiment} presents the experimental results. Secion~\ref{sec:con} concludes this work.
\section{System Model}\label{sec:System_overview}
We consider a real-time distributed platform, with multiple homogeneous single-core CPUs (communication is not considered in this work). Let $\mathcal{E} = \{e_{1},\dots,e_{n}\}$ be the set of CPUs. The functional layer is described by a set of independent tasks $\mathcal{T} = \{\tau_{1},\dots,\tau_{m}\}$. Each task $\tau_{i}$ has a fixed period $t_{i}$, a deadline $d_{i}$, a worst case execution time (WCET) $c_{i}$ and a static priority $p_{i}$. We assume that the system is subject to uncertainties such as external disturbance and transient soft errors. To alleviate the impact of uncertainties, we assume that
\begin{enumerate*}[label={\alph*)},font={\bfseries}]
    \item some tasks can be equipped with error detection and recovery techniques, and
    \item some control tasks can tolerate certain degree of deadline miss.
\end{enumerate*}
In this study, we consider two types of fault-tolerance techniques, EED and EOC. For each task $\tau_i \in \mathcal{T}$, we use a variable $o_{\tau_{i}}$ to denote the choice of fault-tolerance technique. Once a transient soft error is detected, corresponding task re-execution is followed to correct the soft error. Due to the difference between the two fault-tolerance techniques and the random arrival of soft errors, we model each task as a superposition of typical and overload activation, as explained below in details.

\subsection{Error Detection Strategy and Modeling}
For simplicity, we consider a single-error model in this work, where we assume that there is at most one transient soft error within the task set hyper-period (in practice this covers vast majority of the cases). Let $C_{i}$ be the worst case execution time with error detection for task $\tau_{i}$, and $c_{i}$ be the original WCET when error detection is not applied. Then the worst case execution time with error detection for any task $\tau_{i}$ can be defined as in~\cite{zheng2015analysis}:
\begin{equation}\label{eq:WCET_with_error}
C_{i} = c_{i} + \rho_{i}\big(o_{i}(c_{i}+\Lambda_{i})+(1-o_{i})\Delta c_{i}\big),
\end{equation}
where $\rho_{i}$ denotes whether any error detection (EOC or EED) is applied for task $\tau_{i}$, $\Lambda_{i}$ the time for output comparison and $\Delta c_{i}$ the EED overhead. Moreover, $o_{i}=1$ if EOC is selected, otherwise $o_{i} = 0$. 
Note that $C_{i}$ only includes WCET and error detection overhead. Once an error is detected, a re-execution is scheduled immediately. Let $CR_{i}$ be the error recovery/re-execution time for a task $\tau_{i}$. We have
\[
CR_{i} = \rho_{i}\big(c_{i}+(1-o_{i})\Delta c_{i}\big).
\]
As we will discuss later, $CR_{i}$ corresponds to the execution time of an overload activation due to transient soft errors, while $C_{i}$ is the execution time of regular periodic activation.

\subsection{Task Execution Model}
Considering the sporadic nature of transient soft errors, we characterize our task model by its activation pattern and execution time pattern, similarly as in~\cite{leonie2019improving}. Each pattern is further distinguished by a typical component and an overload component. More specifically, for each task with any error detection technique, the periodical activation pattern corresponds to the typical model, whereas the sporadic overload is due to addressing the transient soft errors. They are formally defined below.


\begin{definition}{Event models~\cite{leonie2019improving}: }
The event models $\eta_{i}^{-,(tp)}(\Delta t)$ and $\eta_{i}^{+,(tp)}(\Delta t)$ ($\eta_{i}^{-,(o)}(\Delta t)$ and $  \eta_{i}^{+,(o)}(\Delta t)$, respectively) provide lower and upper bound on the number of typical (overload, respectively) activations of task $\tau_{i}$ during any time interval $[t,t+\Delta t)$.
\end{definition}
Due to the periodicity of task events, we have $\eta_{i}^{-,(tp)} = \eta_{i}^{+,(tp)} = \lceil \frac{\Delta t}{t_{i}} \rceil$. There is a minimal interval $\Delta t_{error}$ between two consecutive soft errors and $\eta_{i}^{-,(o)} = \eta_{i}^{+,(o)} = \lceil \frac{\Delta t}{\Delta t_{error}} \rceil$, if applicable.
For simplicity, we assume that the worst-case event model $\eta_{i}^{+} = \eta_{i}^{+,(tp)} +\eta_{i}^{+,(o)}$.
\begin{definition}{Execution time model~\cite{leonie2019improving}: }
The execution time model $\gamma_{i}^{-,(tp)}(n)$ and  $\gamma_{i}^{+,(tp)}(n)$ ($\gamma_{i}^{-,(o)}(n)$ and $\gamma_{i}^{+,(o)}(n)$, respectively) provide  lower and upper bound on the typical (overload, respectively) share of the service demand required by any $n$ consecutive activation of task $\tau_{i}$.
\end{definition}
In this work, we have $\gamma_{i}^{-,(tp)}(n) = \gamma_{i}^{+,(tp)}(n) = n\times C_{i}$, where $C_{i}$ is the WCET with error detection as defined in~\eqref{eq:WCET_with_error}. Similarly, $\gamma_{i}^{-,(o)}(n) = \gamma_{i}^{+,(o)}(n) = n\times CR_{i}$.

Throughout the paper, we assume that some tasks can miss certain number of deadlines. we employ a general representation to characterize such task timing requirement. Let $\zeta_{i} = \{(k_{i}^{1},N_{i}^{1}),\dots,\allowbreak (k_{i}^{n_{i}},N_{i}^{n_{i}})\}$ be the set of the weakly-hard constraints of task $\tau_{i}$, where $(k_{i}^{j},N_{i}^{j})$ means for any $N_{i}^{j}$ consecutive activations of task $\tau_{i}$, at most $k_{i}^{j}$ deadline misses are allowed. $(0,1)$ is the special case for hard deadline tasks. The system is schedulable if
\begin{equation}
\label{eqn:sched_constraints}
    dmm_{i}(N_{i}^{j}) \leq k_{i}^{j}, \forall j, 1\leq j \leq n_{i}, \forall i,
\end{equation}
where $dmm_{i}(N_{i}^{j})$ is the maximum number of deadline misses of task $\tau_{i}$ in any $N_{i}^{j}$ consecutive activations. We further assume that tasks will continue running until they finish when their deadlines are missed.

\subsection{Control Model}
We consider linear time-invariant (LTI) control tasks. The system dynamic is modeled as:
\begin{gather*}
    \dot{x}(t)=Ax(t)+Bu(t),\\
    y(t)=Cx(t),
\end{gather*}
where $A$, $B$ and $C$ are system matrices, and $x(t)$, $u(t)$ and $y(t)$ are vectors representing the system state, control input and system output at time $t$, respectively. We further assume that the control task is activated periodically and follows the Logical Execution Time (LET) diagram. The LET implementation applies control input at the deadlines and provides fixed closed-loop delay~\cite{pazzaglia2018beyond,Frehse_RTSS_14}. The corresponding discrete-time system dynamics with certain sampling period $h$ is given by ~\cite{Astrom97}:
\begin{equation}\label{eq:discrete_states}
    x[k+1]=A_d x[k]+B_{d,0}u[k]+B_{d,1}u[k-1],
\end{equation}
where $B_{d,0}=\int\limits_0^{h-D} e^{As}\cdot B ds$ and $B_{d,1}=\int\limits_{h-D}^{h} e^{As}\cdot B ds$. $D$ is the relative deadline. By defining an augmented state matrix $z[k] =\begin{bmatrix}x[k] \\ u[k-1]\end{bmatrix}$, we can rewrite the delayed system in (\ref{eq:discrete_states}) as:
\begin{equation}\label{eq:aug_discrete_state}
    z[k+1]=A_{aug}z[k]+B_{aug}u[k],
\end{equation}
where {\small $A_{aug}=\begin{bmatrix}A_d & B_{d,1}\\ \mathbf{0} & \mathbf{0}\end{bmatrix}$, $B_{aug}=\begin{bmatrix}B_{d,0}\\ \mathbf{I}\end{bmatrix}$, $C_{aug}=\begin{bmatrix}C & \mathbf{0}\end{bmatrix}$}, with  $\mathbf{0}$ and $\mathbf{I}$ denoting zero matrix and identity matrix of suitable dimensions, respectively. The control law $u[k]=-Kz[k]$ is calculated by pole place technique~\cite{Kautsky85}.

\vspace{-3pt}
\begin{figure*}
    \centering
    \includegraphics[width=0.85\textwidth]{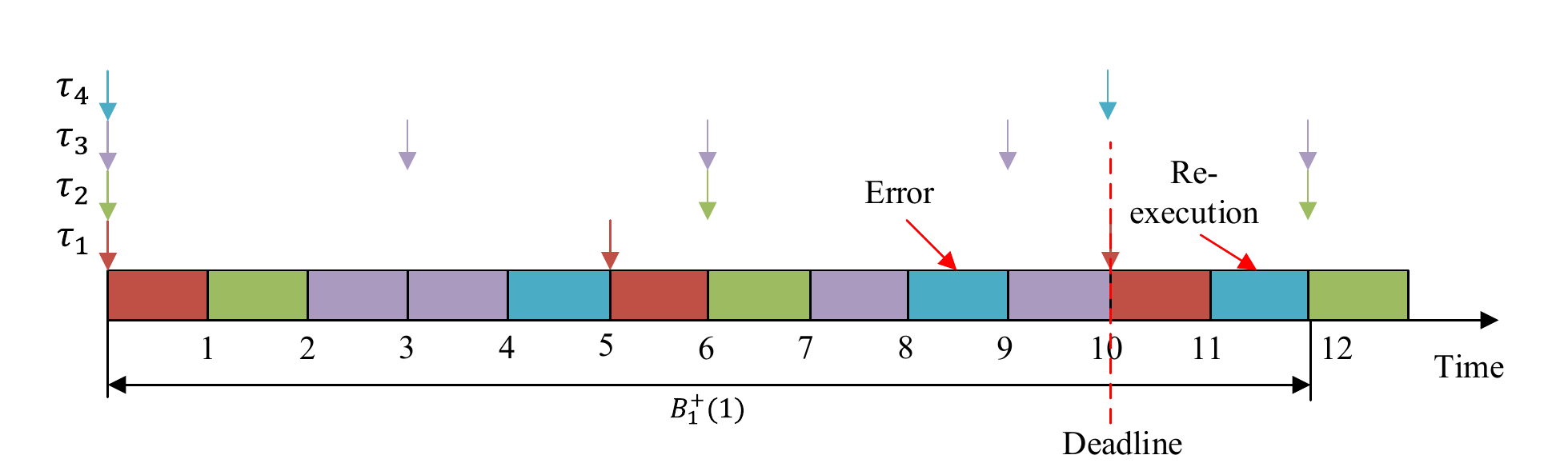}
    \caption{Illustrating example: task $\tau_{4}$ cannot be applied with EOC under hard timing constraints.}
    \label{fig:TWCA}
\end{figure*}

\smallskip
\noindent

\section{Problem Analysis and Formulation}\label{sec:prob_form}


In this section, we introduce our analysis and formulation of the problem, including the definition for a system-level error coverage metric and the analysis for control stability and cost.


\smallskip
\noindent
\textbf{Illustrating Example:} To illustrate how we leverage the weakly-hard constraints, let us consider 4 tasks running on a single-core CPU as defined in Table~\ref{tab:demo_setting} and shown in Figure~\ref{fig:TWCA}. If there is no error detection applied to these tasks, the taskset is schedulable under hard timing constraints. If we want to add EOC to the control task $\tau_4$, the WCET (with error detection) of the control task becomes $2$. The system is still schedulable when no soft error occurs. However, if there is an error, the control task has to schedule a re-execution job and the system with hard timing constraints is no long schedulable (as the re-execution job of the control task will miss its deadline), as shown in Figure~\ref{fig:TWCA}. 

If the control task is robust enough and can tolerate some deadline misses, we can leverage weakly-hard constraints to improve its fault tolerance. For instance, let us assume $\tau_4$ 
satisfies $(2, 10)$ weakly-hard constraint, the system can be proven schedulable with EOC applied to $\tau_4$.
\begin{table}[!h]
\caption{Task set of the illustrating example.}
\begin{tabular}{|c|c|c|c|c|}
\hline
Task name & $\tau_{1}$ & $\tau_{2}$ & $\tau_{3}$ & Controller $\tau_{4}$ \\ \hline
Period    & 5      & 6      & 3      & 10           \\ \hline
WCET      & 1      & 1      & 1      & 1            \\ \hline
\end{tabular}
\label{tab:demo_setting}
\end{table}

\vspace{-3pt}
\subsection{Error Coverage}
We define a system-level \textit{error coverage} metric as the probability that either the transient soft errors are either detected by the error detection technique or the errors occur during the idle time\footnote{For system idle time, transient errors like memory errors may still occur. We assume that the probability that the program is affected by such error is negligible, although we can extend our formulation to cover idle time error.}. For a single-core CPU, assuming $K$ uniformly distributed soft errors can happen within a hyper-period, the error coverage can be approximated as defined in~\cite{zheng2015analysis}:
\[
P \approx \sum_{i = 0}^{K}\sum_{j = 0}^{i}
    \begin{pmatrix}
      K \\
      i
    \end{pmatrix}\cdot
    \begin{pmatrix}
      i \\
      j
    \end{pmatrix}\cdot
  (\frac{\alpha t_{eed}}{T_{hyper}})^{j}\cdot(\frac{\beta t_{eoc}}{T_{hyper}})^{i-j}\cdot(\frac{t_{idle}}{T_{hyper}})^{K-i},
\]
where $\alpha$ and $\beta$ are the average probabilities that an error is detected by EED and EOC, respectively. Here, $t_{eed}$, $t_{eoc}$ and $t_{none}$ are the time spent by tasks using EED, EOC and no error detection, respectively. $T_{hyper} = t_{eed} + t_{eoc} + t_{none} + t_{idle}$. For our study, we assume $K = 1$ and the above equation can be further approximated as follows according to~\cite{zheng2015analysis}:
\[
 P \approx 1 - \frac{\sum_{\tau_{i}\in \mathcal{T}} (1-\varepsilon_{\tau_{i}})C_{i}}{t_{i}},
\]
where $\varepsilon_{\tau_{i}}$ is the error detection rate for task $\tau_{i}$. In our experiment, we set $\alpha = 0.7$ and $\beta = 1$, similarly as in~\cite{gao2013using, zheng2015analysis}. In our work, we assume that there is an error coverage requirement $EC\_Threshold$ defined, such that $P \leq EC\_Threshold$.


\subsection{Control Stability and Cost}\label{sec:stability_control_cost}
We consider stabilization controller that can bring the system back to the equilibrium state after a disturbance. Moreover, due to potential deadline misses, some control inputs may not always be applied on time. Following the LTE diagram, we assume that if a control task misses its deadline, the last control input will be used. The control cost is defined as the number of sampling periods needed to bring the system back to the equilibrium state~\cite{wang2020crosslayer}. The  control input delay at the $k$-th instance can be bounded by:
\begin{equation*}
    \psi_{k} \leq \psi_{max} = \lceil \frac{r_{i}}{t_{i}} \rceil,
\end{equation*}
here $r_{i}$ is the worst-case response time obtained by the schedulability analysis (detailed later in Section~\ref{subsec:schedulability}). Under such deadline miss case, the system state can be captured by the augmented state vector:
\[
\xi[k] = [x^{T}[k],u^{T}[k-1],\dots,u^{T}[k-\psi_{max}]]^{T}
\]
and the system dynamic can be re-written as:
\begin{gather}
    \xi[k+1] = A_{\xi}\xi[k] + B_{\xi}u[k]\\
    A_{\xi}[k] = \begin{bmatrix}
    A_{d} & B_{1} & \dots & B_{\psi_{max}-1} & B_{\psi_{max}}\\
    0 & I & \dots & 0 & 0\\
     & & \ddots & & \\
     0 & 0 & \dots & I & 0
    \end{bmatrix},
    B_{\xi} = \begin{bmatrix}
    0\\
    I\\
    0\\
    \vdots\\
    0
    \end{bmatrix}
\end{gather}
where $B_{\psi_{k}}=B_{d,1}$ and $B_{i}=0,\forall i \neq \psi_{k}$. $u[k-\psi_{k}]$ is the latest control input. The above system dynamic can be simplified as $\xi[k+1] = (A_{\xi}[k]-B_{\xi}[k]K_{\xi})\xi[k] = \phi[k]\xi[k]$, where $K_{\xi} = [K,\textbf{0}]$.

\smallskip
\noindent
\textbf{Control Stability:} 
Assuming we are given the deadline hit/miss pattern of a control task within a hyper-period, thus the delay $\psi_{k}$ in a hyper-period and the transition matrix $A_{\xi}[k]$ of each control period are known ($\forall k \in [0,N]$). Thus, we have:
\begin{align*}
    \xi[k+N] &= \phi[k+N-1]\dots\phi[k+1]\xi[k]\\
    &=\prod_{i = k+N-1}^{0}\phi[i]\prod_{j=N-1}^{k}\phi[j]\xi[k]\\
    &=\Phi_{k}\xi[k]
\end{align*}
The above system under deadline misses is asymptotically stable if the eigenvalues of $\Phi_{k}$ are within the unit circle for all $k$~\cite{wang2020crosslayer}.

\smallskip
\noindent
\textbf{Control Cost:} 
We define the cost of a control task as its ability to reject an external disturbance. Formally, let us assume that an external disturbance occurs at the $k$-th job and brings the system state to $x[k]$. We consider the disturbance is rejected if the residual disturbance after $r$ sampling period satisfies:
\[
\frac{||x[k+r]||}{||x[k]||} \leq J_{th}, \quad \forall r \geq h_{k},
\]
where $J_{th}$ is a pre-defined threshold. Let $\xi[k+r]=\Phi_{k+r,k}\xi[k]$. Then $x[k+r]$ can be expressed as:
\[
x[k+r] = \hat{I}\xi[k+r]=\hat{I}\Phi_{k+r,k}\xi[k]=\hat{I}^{T}\Phi_{k+r,k}\hat{I}x[k],
\]
where $\hat{I} = \begin{bmatrix}I_{n\times n} & 0_{n\times m}\end{bmatrix}^{T}$.
Then, the control cost is defined as:
\begin{definition}
The control cost metric of a control task $\tau_{i}$ is defined as $J_{i} = \max\{h_{k}|\forall k\}$, where $h_{k}$ satisfies:
\begin{equation}\label{eq:individual_control_perf}
||\hat{I}^{T}\Phi_{k+r,k}\hat{I}|| \leq \frac{||x[k+r]||}{||x[k]||} \leq J_{th}, \quad \forall r \geq h_{k}.
\end{equation}
\end{definition}

\begin{figure}[!htbp]
    \vspace{-2pt}
    \centering
    \includegraphics[width=0.75\columnwidth]{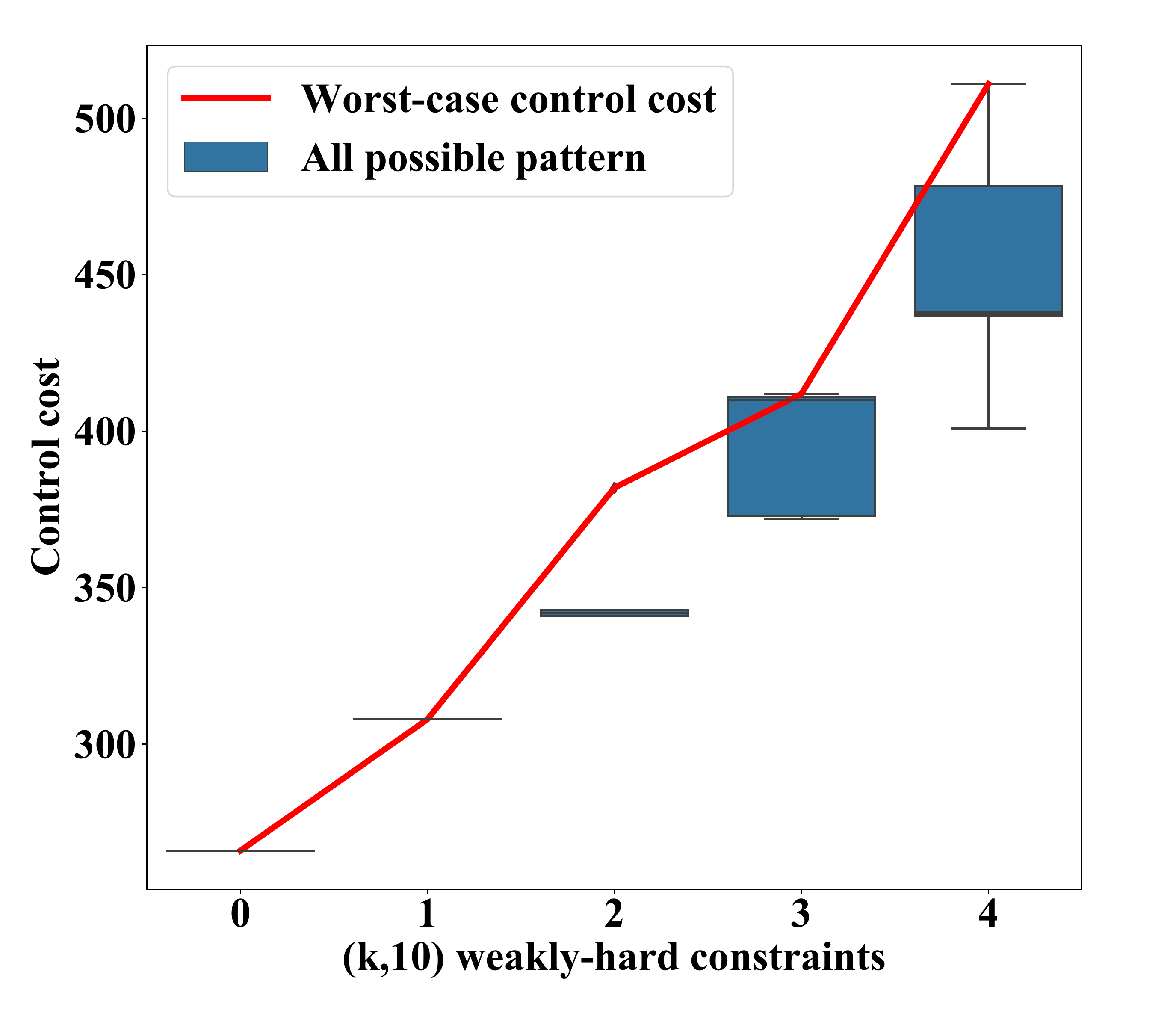}
    \caption{Control cost approximation of a demo controller, where $k$ ranges from $0$ to $4$, when $n=10$.}
    \label{fig:control_approx}
\end{figure}

\smallskip
\noindent
\textbf{Approximation of Control Cost:}  
Calculating the precise control cost through event-based simulation is time-consuming. Thus, we try to approximate the control cost by utilizing the deadline miss bound provided by the schedulability analysis (detailed in Section~\ref{subsec:schedulability}).
Assuming that a control task satisfies the $(k,n)$ weakly-hard constraint, where $k$ is an upper-bound of the number of deadline misses in $n$ consecutive activations. 
Then, for any $i \in \mathbb{N}$, $A_{\xi}[i]$ can take at most $k$ different forms. Similarly, $\phi[i]$ can take at most $k$ different forms.
Since $k$ and $n$ are typically small numbers, we can exhaustively search all patterns of length $n$ that satisfy the $(k,n)$ deadline miss bound and find out the worst-case pattern as our approximation. 
Figure~\ref{fig:control_approx} shows result of a demo cruise control controller~\cite{Goswami_TCST14}, when $n=10$ and $k$ changes from $0$ to $4$. The red line is the approximated control cost\footnote{In this demo example, for $k=2$, there is an outlier point when the deadline misses are evenly distributed.}.

\subsection{Overall Problem Formulation}
Our optimization objective is the overall system-level control cost defined as the weighted sum of each individual control cost:
\begin{equation}\label{eq:system_perf}
    \mathcal{J} = \sum_{\tau_{i}\in\mathcal{T}_{J}} \omega_{i}\frac{J_{i}}{J^{des}_{i}},
\end{equation}
where $\omega_{i}$ are the given weights, $\mathcal{T}_{J}$ is the set of control tasks, and $J^{des}_{i}$ is the desired control cost of task $\tau_{i}$ assuming no deadline miss occurs. We can formulate our problem as:

\smallskip
\noindent
\textbf{(P1)} Given $\mathcal{E},\mathcal{T},\mathcal{T}_{J}$, optimize task assignment, selection of error detection mechanism $\mathcal{O} =\{o_{\tau_{1}},\dots,o_{\tau_{|\mathcal{T}_{J}|}}\}$ such that
\begin{myitemize}
    \item schedulability constraints as defined in Equation~\eqref{eqn:sched_constraints} are satisfied,
    \item the stability of control tasks as defined above are satisfied, and
    \item the error coverage requirement $EC\_Threshold$ is satisfied.
\end{myitemize}

\section{Meta-heuristic Algorithm for Design Space Exploration}\label{sec:meta_heuristic}

Our meta-heuristic algorithm relies on the schedulability analysis under weakly-hard constraints and the fault-tolerance model. In this section, we first introduce the two schedulability analysis methods and then we present our optimization algorithm.

\subsection{Schedulability Analysis}\label{subsec:schedulability}
We assume that tasks running on the same CPU is scheduled by the static priority preemptive (SPP) scheduling policy. In this study, we leverage two different schedulability analysis methods for weakly-hard systems. One extends the work from~\cite{Xu_ECRTS_15}, where the deadline miss model can be upper-bounded by employing typical worst-case analysis (TWCA). The other extends the event-based schedulability analysis in~\cite{liang2019security}. The event-based schedulability simulates the execution of tasks within a hyper-period and derives the deadline miss patterns for all tasks in a single run. In the following, we will briefly introduce TWCA and then a detailed explanation of the event-based simulation with error injection.

\subsubsection{Bounding Deadline Miss Model Using TWCA}
Our schedulability analysis extends the ideas from ~\cite{Xu_ECRTS_15,leonie2019improving}. The state-of-art technique ~\cite{Xu_ECRTS_15} is an improved version of ~\cite{Quinton_DATE_12}. In~\cite{Quinton_DATE_12,Xu_ECRTS_15}, the task activation model is a superposition of typical activation and sporadic overload. The typical activation is assumed to be feasible whereas the overload activations can cause at most $m$ deadline misses out of $k$ consecutive activation of a task. However, neither ~\cite{Quinton_DATE_12} nor \cite{Xu_ECRTS_15} distinguishes the execution time of different activation classes (i.e. regardless of whether typical or overload activation). In~\cite{leonie2019improving}, the authors extend the task model with sporadic long execution time overload and the TWCA algorithm is extended based on the result in~\cite{Quinton_DATE_12}. In this study, we borrow the task model in~\cite{leonie2019improving} and bound the deadline misses by counting the number of possible overload activations with the consideration of fault-tolerance techniques.

\smallskip
\noindent
\textbf{Worst-case Response Time of Typical Activation:} Due to error detection techniques considered in this work, our approach to calculate the worst-case response time is different compared to the common practice as in~\cite{Xu_ECRTS_15,leonie2019improving,Quinton_DATE_12}. More specifically, we explicitly consider the potentially varying execution time of recovery jobs. Let $B_{i}^{+}(q)$ denotes the maximum time needed to process $q$ typical activations of task $\tau_{i}$ within any busy window, where the transient soft error may occur\footnote{We assume that the time interval between two consecutive soft errors  are large enough such that soft error will not happen during the execution of recovery jobs.}:
\begin{align}
    B_{i}^{+}(q) =& \gamma_{i}^{+,(tp)}(q) + \sum_{\tau_{j} \in hp(\tau_{i})} \gamma_{j}^{+,(tp)}(\eta_{j}^{+,(tp)}(B_{i}^{+}(q)))\\
    & +max\{\gamma_{j}^{+,(o)}(\eta_{j}^{+,(o)}(B_{i}^{+}(q)))|\forall \tau_{j} \in hp(\tau_{i})\}.
\end{align}
Here, the first term is the service demand of the $q$ typical activations; the second term is the interference from higher priority tasks; and the last term is the maximum possible overload service demand due to transient soft error.

\begin{definition}{Worst-case level-i busy window~\cite{Xu_ECRTS_15}:}
A worst-case level-i busy window, denoted as $BW_{i}$, is the maximal time window during which tasks of equal or higher priority than task $\tau_{i}$ have pending jobs. 
\end{definition}
$BW_{i}$ can be calculated as following:
\begin{equation}
    BW_{i} = B_{i}^{+}(K_{i}),
\end{equation}
where
\begin{align*}
    K_{i} &= \min\{q \geq 1: B_{i}^{+}(q) \leq \eta_{i}^{-,(tp)}(q+1)\}.
\end{align*}
The worst-case response time can be calculated as:
\[
r_{i} = \max_{1\leq q\leq K_{i}}\{B_{i}^{+}(q)-\delta_{i}^{-,(tp)}(q)\},
\]
where $\delta_{i}^{-,(tp)}$ is the event distance function of typical activations.

TWCA assumes that task $\tau_{i}$ is schedulable in the typical model (i.e., no transient soft error). However, in the worst case, out of the $K_{i}$ activations in the worst-case busy window, some may miss their deadlines. Let us denote these deadline misses by $N_{i}$ and thus $N_{i} = \{q \in \mathbb{N}| 1\leq q \leq K_{i} \land B_{i}^{+}(q)-\delta_{i}^{-,(tp)}(q) > d_{i}\}$. These deadline misses are caused by the overload activations due to transient soft errors. Thus, in order to bound the maximum number of deadline misses, we just need to find how many recovery jobs may affect the $k$ consecutive typical activations.

The maximum time window $\Delta T^{j,i}_{k}$ during which the overload activation of $\tau_{j}$ may impact the $k$ activation of task $\tau_{i}$ can be calculated by~\cite{leonie2019improving}:
\[
\Delta T^{j,i}_{k} = BW_{i}+\delta^{+,(tp)}_{i}(k)+r_{i}.
\]

Finally, the number of deadline misses is then bounded by:
\begin{equation}
    dmm_{i}(k) = |N_{i}|\times \ceil[\bigg]{\frac{\Delta T^{j,i}_{k}}{\Delta t_{error}}}.
\end{equation}

\subsubsection{Event-based Simulation for Exact Deadline Miss Pattern}
The aforementioned schedulability analysis only guarantees a pessimistic upper-bound to the number of deadline misses within $k$ consecutive activations. However, the exact deadline hit/miss patterns sometimes have a non-negligible effect on the control cost. Moreover, due to the randomness of the soft errors, the activation patterns of recovery jobs are not clear. Thus, we build an event-based simulation with error injection. 
The pseudo-code is shown in Algorithm~\ref{al:event_sim}. 

Our event-based simulation records the time-stamp of each event such as the job release time, finish time, etc. we denote the $j$-th invocation of task $\tau_i$ as job $\theta_{ij} = (s_{\theta_{ij}}, c_{\theta_{ij}})$, where $s_{\theta_{ij}} = j\cdot t_{\tau_i}$ is the release time of the job. $c_{\theta_{ij}}$ keeps track of the remaining computation time of the job. 
For each task $\tau_i$, we record its deadline miss patterns in an array $Miss[i]$, where $Miss[i][j]=true$ if $\tau_i$'s $j$-th job $\theta_{ij}$ misses its deadline. 
$event\_queue$ and $job\_queue$ are two job priority queues to store the unreleased jobs and pending jobs, respectively. $event\_queue$ is sorted by the job release time $s_{\theta_{ij}}$ while $job\_queue$ is sorted by the task priority. 

We assume that the soft error can arrive in any time during the hyper-period. The algorithm first pushes all typical activations into the $event\_queue$. Function $InjectError()$ will try to inject a soft error for each event and the corresponding re-execution job will be pushed into $event\_queue$. Then we conduct an efficient even-based simulation of the whole hyper-period (lines 10-28). Note that if the event belongs to a task without error detection technique, we just skip to the next event. For a time point $cur\_time$, any jobs that can be released are popped from the $event\_queue$ and then pushed into the $job\_queue$, and the highest priority job in the $job\_queue$ is scheduled to run. 
Here, $\theta_{kl}$ is the scheduled job at $cur\_time$ and $\theta_{ij}$ is the next job to release. 
Then, the simulation moves to the next time point. 

If the scheduled job has not finished at time $next$, it will update its remaining execution time $c_{\theta_{ij}}$ and be pushed back to the $job\_queue$.  Every time a job $\theta_{kl}$ finishes, the simulation records whether it misses its deadline in $Miss[k][l]$. After the simulation completes, the function $VerifyWHConstraint()$ counts the maximum deadline misses of any consecutive $N_{i}^{j}$ activations (i.e., $dmm_i(N_{i}^{j})$) to verify whether tasks have met corresponding weakly-hard constraints. Once we have finished the simulation of the whole hyper-period, we clear the error and move to next possible event. The return value indicates whether current system configuration is schedulable under the worst-case soft error scenario.

\begin{algorithm}[!h]
\caption{$EventSim$: Event-based Simulation with Error Injection}
\label{al:event_sim}

\begin{algorithmic}[1]
\STATE $WCRTAnalysis(\mathcal{T})$
\IF {$\forall \tau_{i} \in \mathcal{T}$, $r_{\tau_{i}} \leq d_{\tau_{i}}$}
\RETURN \TRUE
\ENDIF
\FOR {task $\tau_{i} \in \mathcal{T}$}
    \FOR {$j  \in \{0,\dots,\frac{HyperPeriod}{t_{i}}-1\}$}
        \STATE $event\_queue.push(\theta_{ij})$
    \ENDFOR
\ENDFOR
\FOR {e $\in\ event\_queue$}
\IF{$InjectError(e)$}
    \STATE $\theta_{ij} = event\_queue.pop()$,\quad $cur\_time = s_{\theta_{ij}}$
    \WHILE{$cur\_time \leq HyperPeriod$}
    \WHILE{$s_{\theta_{ij}} \leq cur\_time$}
        \STATE $\theta_{ij} = event\_queue.pop()$
        \STATE $job\_queue.push(\theta_{ij})$
    \ENDWHILE
    \IF{$job\_queue$ is empty}
    \STATE $cur\_time = s_{\theta_{ij}}$
    \ELSE
    \STATE $\theta_{kl} = job\_queue.pop()$,\quad $next = s_{\theta_{ij}}$
    \STATE $response = cur\_time + c_{\theta_{kl}}$
    \IF{$response \leq next$}
    \IF {$response \leq s_{\theta_{kl}} + d_{\tau_{k}}$}
    \STATE $Miss[k][l] =$ \FALSE
    \STATE $cur\_time = response$
    \ELSE
    \STATE $Miss[k][l] =$ \TRUE
    \STATE $cur\_time =$ $\max$($s_{\theta_{kl}} + d_{\tau_{k}}$, $cur\_time$)
    \ENDIF
    \ELSE
    \STATE $c_{\theta_{kl}} = c_{\theta_{kl}} - (next - cur\_time)$
    \STATE $job\_queue.push(\theta_{kl})$, $cur\_time = next$
    \ENDIF
    \ENDIF
    \ENDWHILE
    \STATE $ClearError(e)$
    \STATE $schedulability = VerifyWHConstraint(Miss)$
    \IF{$!schedulability$}
        \RETURN \FALSE
    \ENDIF
\ENDIF
\ENDFOR
\RETURN \TRUE
\end{algorithmic}
\end{algorithm}

\subsection{Optimization Algorithm}
In this section, we develop a meta-heuristic that tries to optimize control cost  while meeting various constraints. We first use a simple heuristic to decide the initial choice of error detection technique for each task. Then an initial solution for the whole system is generated by using a bin-packing scheme~\cite{coffman1984approximation}. Finally, we use simulated annealing (SA) to further explore the design space.

\subsubsection{Initial Solution}
The initial system configuration is generated by two steps. In the first step, we sort the tasks based on their utilizations in an ascending order. Then, starting from the task with the lowest utilization, we assign EED to each task until the error coverage requirement is met. After the decision for error detection technique is made, we map the taskset onto the underlying hardware platform using a bin-packing algorithm. The priority of each task is then assigned using the deadline monotonic priority assignment. The pseudo-code for obtaining the initial solution is shown in Algorithm~\ref{alg:initial}.
\begin{algorithm}[htbp]
\caption{Obtaining Initial Solution}
\label{alg:initial}
\begin{algorithmic}[1]
\SetAlgoNoLine
\REQUIRE \text{taskset $\mathcal{T}$, error coverage requirement $EC\_Threshold$}
\STATE $\mathcal{T}_{sort} = utilSort()$
\STATE $p = getEC()$
\WHILE{$p < EC\_Threshold$}
    \FOR {$\tau_{i} \in \mathcal{T}_{sort}$}
        \STATE $\tau_{i}.o_{\tau_{i}} = max(1,tau_{i}.o_{\tau_{i}}+1)$
        \STATE $p = getEC()$
    \ENDFOR
\ENDWHILE
\RETURN $binPacking(\mathcal{T}_{sort})$
\end{algorithmic}
\end{algorithm}

\subsubsection{Overall Meta-heuristic with SA}
Algorithm~\ref{alg:sa} shows our meta-heuristic optimization. First, the function $ObtainInitialSolution()$ generates the initial solution as in Algorithm~\ref{alg:initial}. Then, the function $SchedulabilityAnalysis()$ checks system schedulability by using either 
\begin{enumerate*}[label={\alph*)},font={\bfseries}]
    \item the TWCA analysis to give an upper-bound to the deadline miss number of each task and check whether it meets the weakly-hard constraints or
    \item the event-based simulation to obtain the exact deadline hit/miss pattern.
\end{enumerate*}
Then $CalculateCost()$ returns the control cost:
\begin{enumerate*}[label={\alph*)},font={\bfseries}]
    \item if TWCA is used, $CalculateCost()$ gets current control cost by approximation as discussed in Section~\ref{sec:stability_control_cost}, or
    \item if event-based simulation is used,  $CalculateCost()$ calculates the exact control cost using the obtained deadline hit/miss pattern.
\end{enumerate*} 
Function $AddPenalty()$ adds a penalty to the cost value if current solution is unschedulable or the error coverage is below the error coverage requirement $EC\_Threshold$. During each step of the simulated annealing, $S_{cur}$ will randomly move to another configuration $S_{new}$ by either swapping the priority of two tasks or changing the error detection technique or the allocation of a task. If the new configuration cannot be guaranteed to be schedulable under the schedulability analysis, a penalty will be added to the cost value. The new configuration will be accepted if it has a better objective value; otherwise, the acceptance probability will be calculated based on the current temperature and the objective difference.

\begin{algorithm}[htbp]
\caption{Meta-heuristic Optimization with SA}
\label{alg:sa}
\begin{algorithmic}[1]
\SetAlgoNoLine
\STATE $S_{0}=ObtainInitialSolution()$
\STATE $S_{best} = S_{cur} = S_{new} = S_{0}$
\STATE $is\_sched = SchedulabilityAnalysis(S_{0})$ 
\STATE $current\_cost = CalculateCost()$
\STATE $current\_cost = AddPenalty()$
\STATE $\eta_{best} = \eta_{cur} = \eta_{new} = current\_cost$
\WHILE{$T > T^{*}$}
\STATE $k=1$
\WHILE{$k \leq iter\_max$}
\STATE $S_{new} = RandomMove(S_{cur})$
\STATE $\eta_{new} = CalculateCost() + AddPenalty()$
\IF{$\eta_{new} < \eta_{cur}$}
\STATE $S_{cur} = S_{new}, \eta_{cur} = \eta_{new}$
\IF{$S_{new}.is\_sched  == \TRUE \land S_{new}.EC \geq EC\_Threshold \land \eta_{cur} < \eta_{best}$}
\STATE $S_{best} =min(S_{cur},S_{best})$
\STATE $\eta_{best} =min(\eta_{cur},\eta_{best})$
\ENDIF
\ELSIF{$AccepProb(\eta_{new}-\eta_{cur},T) > rand()$}
\STATE $S_{cur} = S_{new}, \eta_{cur} = \eta_{new}$
\ENDIF
\STATE $k = k+1$
\ENDWHILE
\STATE {$T=T*cooling\_factor$}
\ENDWHILE
\RETURN $S_{best}$, $\eta_{best}$
\renewcommand{\algorithmicrequire}{\textbf{Function}}
\end{algorithmic}
\end{algorithm}
\section{Experimental Results}\label{sec:experiment}
We evaluate our proposed approach with an industrial case study and a set of synthetic examples. Our controller tasks are derived based on 4 example LTI systems~\cite{Goswami_TCST14,messner1999control}. The weakly-hard constraints for them are chosen such that the control stability of each task is guaranteed based on the analysis in Section~\ref{sec:prob_form}. Each non-control task is randomly assigned with a $(k,n)$ constraint where $k$ ranges from $0$ to $4$ and $n$ ranges from $10$ to $20$. All experiments are conducted on a server with Intel Xeon Gold 6130 CPU at 2.1 GHz.


\subsection{Synthetic Examples}
We conduct experiments with a set of 50 synthetic examples. Each synthetic example consists of 4 non-control tasks and 4 control tasks, all mapped onto a single-core CPU.

\begin{figure}
    \centering
    \includegraphics[width=0.85\columnwidth]{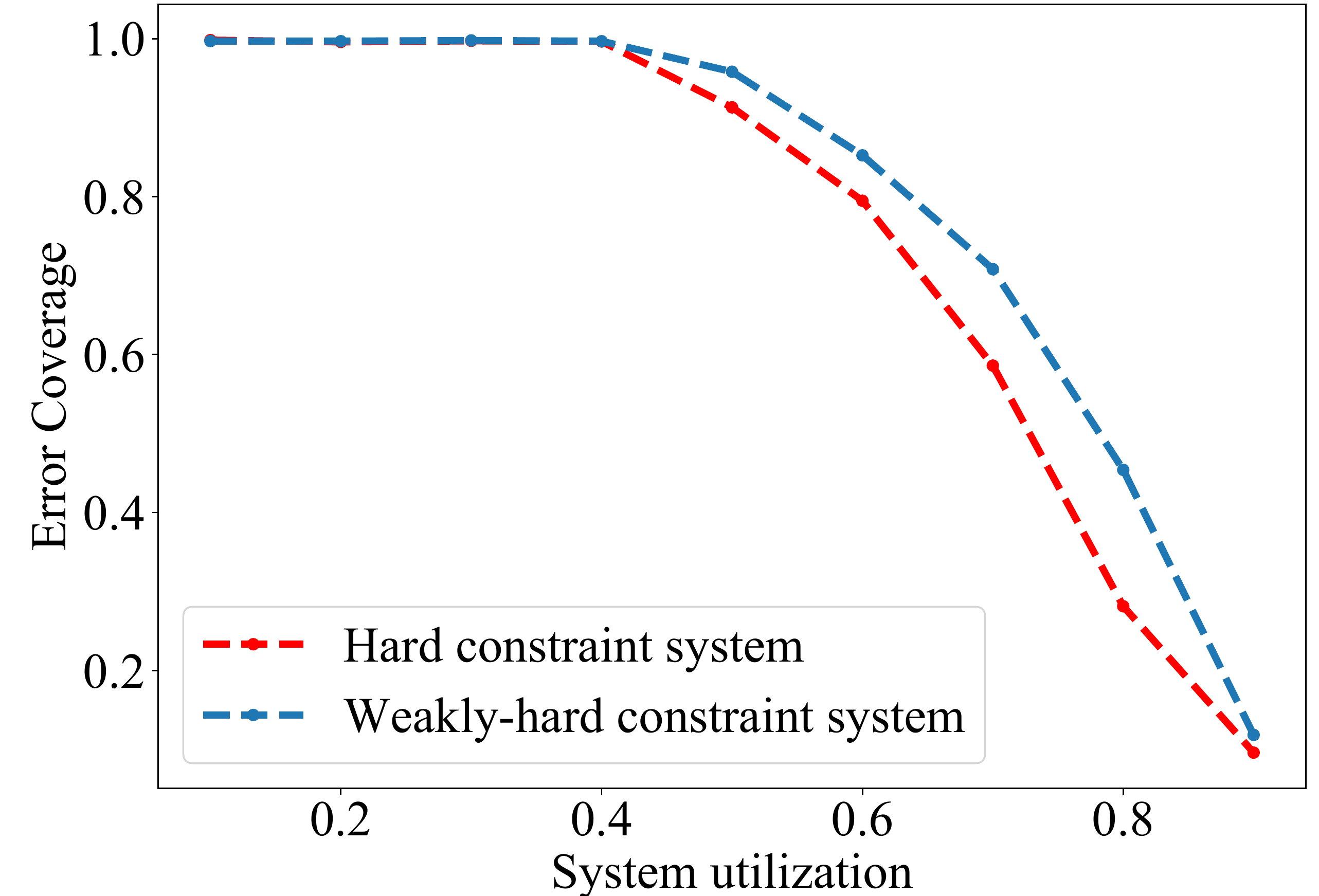}
    \caption{Comparison of average error coverage for weakly-hard-constraint systems and hard-constraint systems.}
    \label{fig:hd_vs_wh}
\end{figure}

\smallskip
\noindent
\textbf{Hard Constraints vs. Weakly-hard Constraints:} To see how much improvement can be obtained from leveraging weakly-hard constraints, we use the event-based SA to explore the maximum error coverage. The average maximum error coverage over the 50 synthetic examples is shown in Figure~\ref{fig:hd_vs_wh}. Both hard-constraint and weakly-hard-constraint systems can achieve $100\%$ error coverage when the system utilization is below 0.4. As the utilization increases, our approach can make use of the scheduling slack obtained from the weakly-hard constraints, i.e., allowing certain tasks to miss their deadlines can enhance the systems fault-tolerance capability. When utilization is 0.9, both weakly-hard and hard-constraint systems are not able to achieve meaningful error coverage.

\begin{figure}[!h]
    \centering
    \includegraphics[width=0.85\columnwidth]{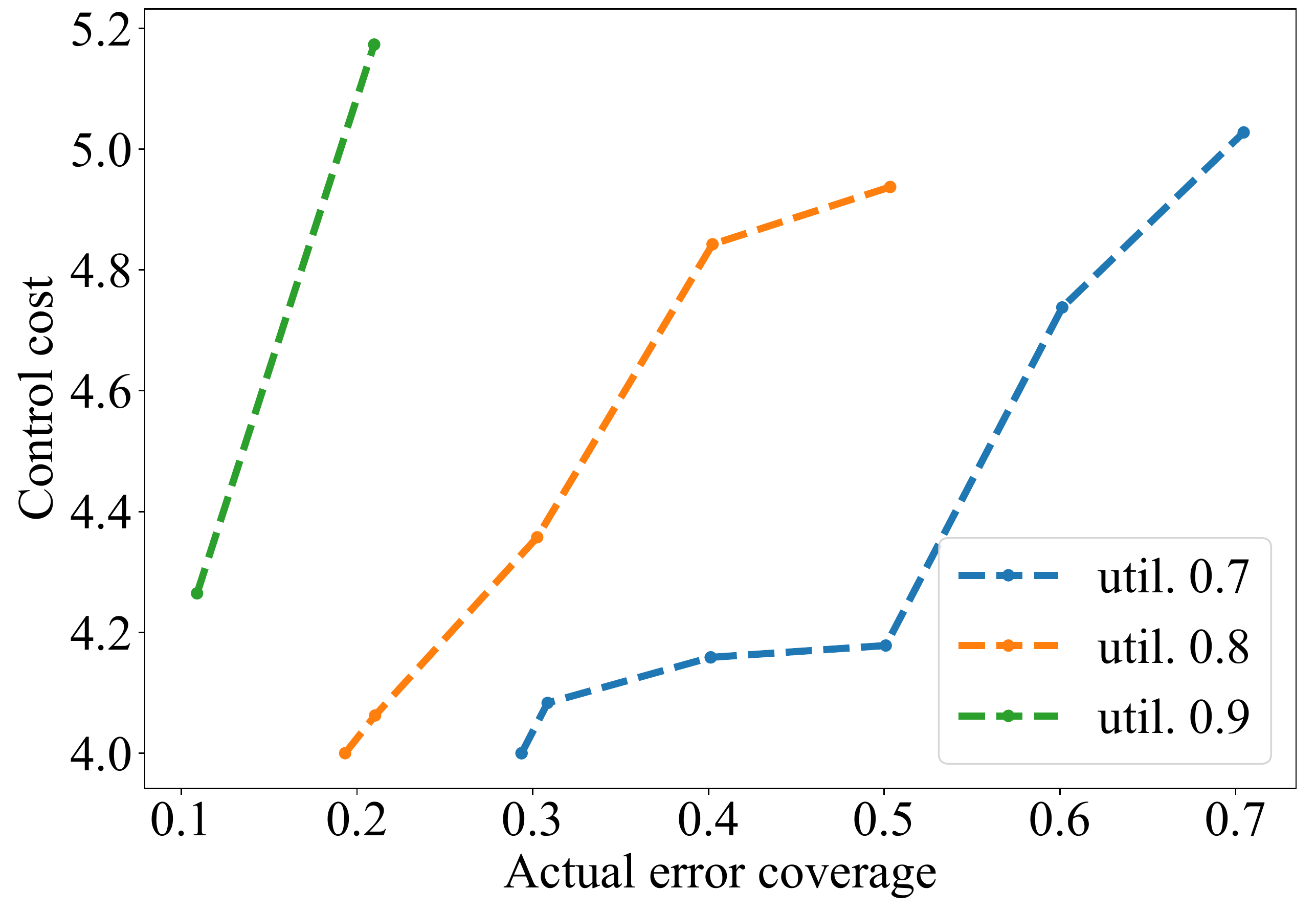}
    \caption{Control cost of different system utilization when error coverage requirement changes from 0.1 to 0.7}
    \label{fig:varying_util}
\end{figure}

\smallskip
\noindent
\textbf{Impact of System Utilization:} Then, we study how the system control cost can be affected by the system utilization and error coverage requirement. Figure~\ref{fig:varying_util} shows the control cost of different system utilization while the actual error coverage increases from 0.1 to 0.7. As expected, when system utilization is 0.9, we can hardly improve the maximum error coverage and the control cost can increase dramatically even though we just add the error coverage requirement by 0.1. For system utilization of 0.7 and 0.8, we are able to find a solution for most of the cases. The maximum error coverage that can be achieved by 0.8 system utilization is around 0.5, while the maximum error coverage of 0.7 system utilization is around 0.7. This information can facilitate the design choices under different system utilizations.

\begin{figure}[!htbp]
    \centering
    \includegraphics[width=0.85\columnwidth]{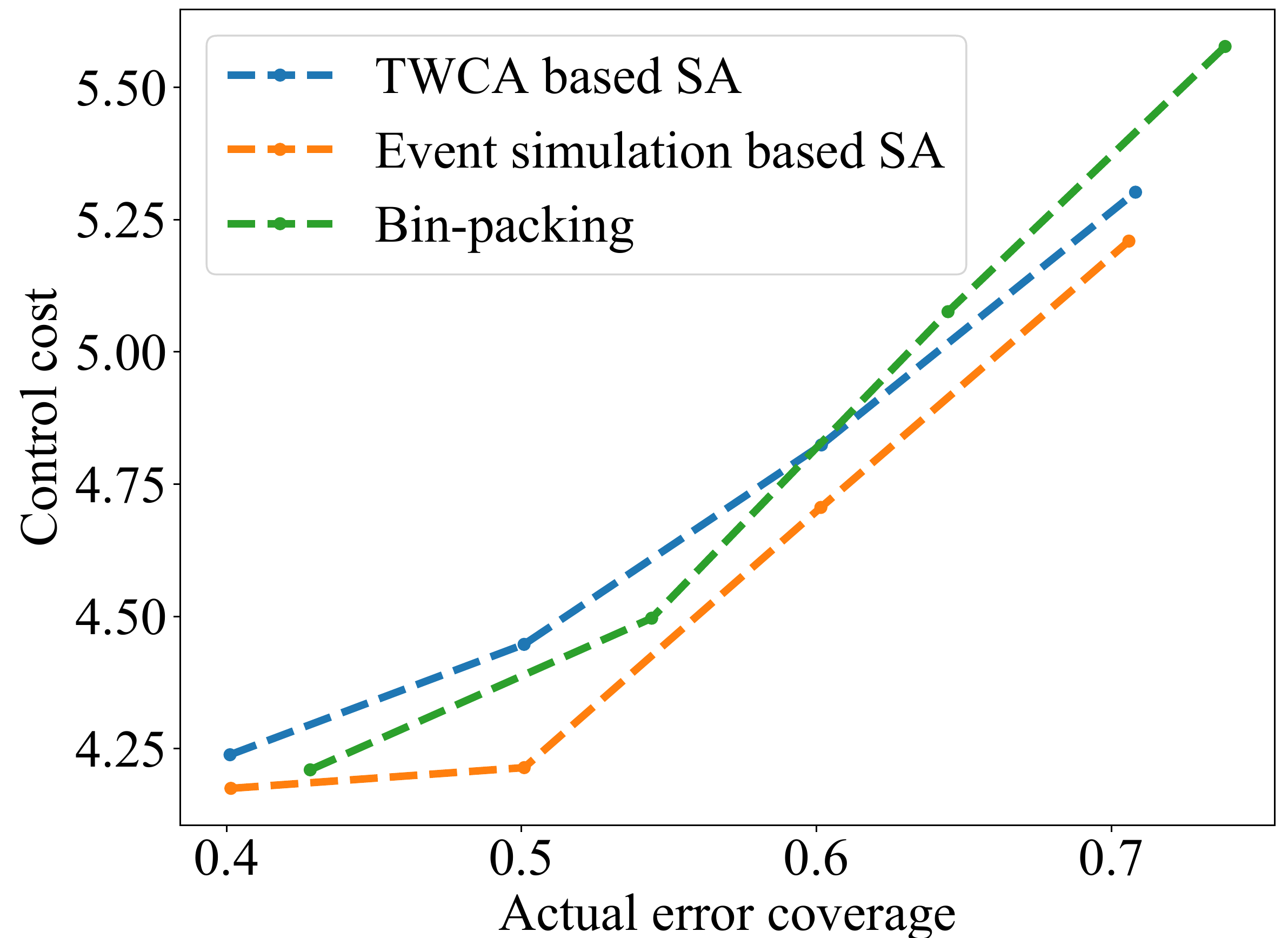}
    \caption{Control cost obtained by different heuristics.}
    \label{fig:alg_comp}
\end{figure}

\smallskip
\noindent
\textbf{Comparison of Heuristic Algorithms:} We also compare the effectiveness of different heuristic algorithms, i.e., the initial solution (bin-packing), the event simulation based simulated annealing, and the TWCA based simulated annealing. We run the three algorithms on a set of synthetic examples with 0.7 system utilization, with error coverage requirement increases from 0.4 to 0.7. Again, notice that the x-axis in Figure~\ref{fig:alg_comp} is the actual error coverage. As we can see, among the three heuristic algorithms, event simulation based SA can output the best solution as it provides the lowest control cost under different error coverage requirements. While bin-packing based approach meets the error coverage requirement, it cannot optimize the control cost.


\subsection{Industrial Case Study: WATERS Challenge}
Our industrial case is derived from the WATERS 2019 Challenge~\cite{hamann124waters}, which consists of 9 tasks and covers a prototype of an advanced driver-assistance system (ADAS). The underlying reference platform is NVIDIA Jetson TX-2 consisting of 6 heterogeneous cores and an integrated GPU. \cite{casini2019addressing} provides a detailed discussion of task modeling and response time analysis, and shows that the original taskset as presented in WATERS 2019 Challenge is unschedulable.

For the purpose of our study, we assume a homogeneous platform and that all tasks are running on ARMv8 A57 cores. 
To make to the taskset schedulable, we scale the WCET of each task by a scaling factor. For our study, we also add four additional control tasks. 
Table~\ref{tab:indus_vary_cpu} shows the maximum error coverage  when the scaling factor changes from $0.3$ to $0.7$, and the number of CPUs changes from $3$ to $5$. We can see that lower utilization and more number of CPUs lead to better error coverage. The error coverage saturates when the scaling factor is 0.3 and 5 CPUs are used.
\begin{table}[!htbp]
\caption{Error Coverage under different scaling factor and number of CPUs.}
\begin{tabular}{|c|c|c|c|c|c|}
\hline
\diagbox{CPU number}{Scaling factor}& 0.3  & 0.4  & 0.5  & 0.6  & 0.7  \\ \hline
3 & 0.68 & 0.3  & n.a. & n.a. & n.a. \\ \hline
4 & 0.85 & 0.67 & 0.46 & 0.15 & n.a. \\ \hline
5 & 1.0  & 0.86 & 0.76 & 0.56 & 0.14 \\ \hline
\end{tabular}
\label{tab:indus_vary_cpu}
\end{table}

\begin{figure}[!h]
    \centering
    \includegraphics[width = 0.85\columnwidth]{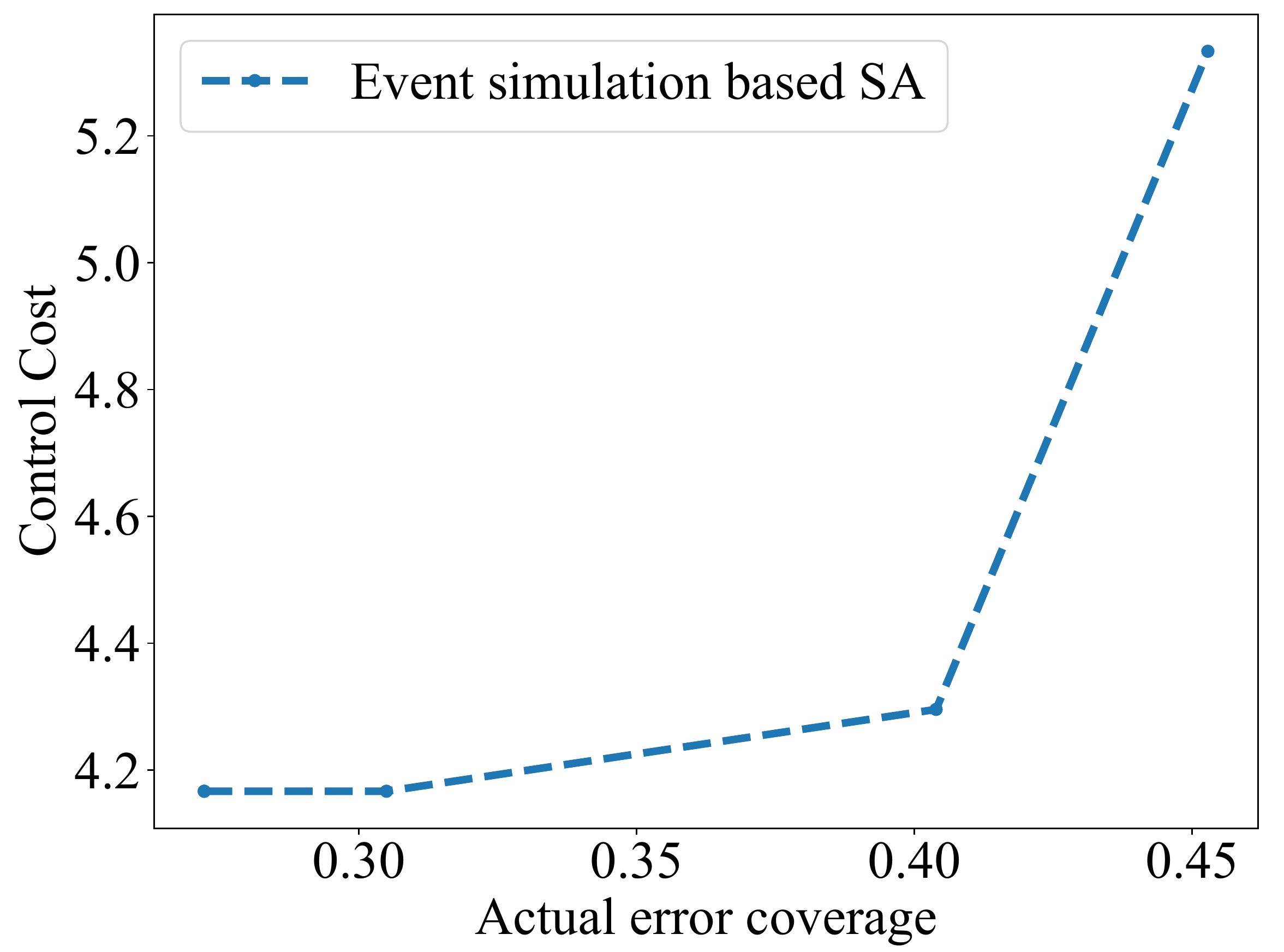}
    \caption{Tradeoff analysis between error coverage and control cost for the WATERS 2019 Challenge example enabled by our approach.}
    \label{fig:waters_2019}
\end{figure}

Figure~\ref{fig:waters_2019} shows the trade-off between error coverage and control cost when the scaling factor is $0.5$ and the number of CPUs is 4. During the experiments, we increase the error coverage requirement changes from 0.1 to 0.45. The x-axis in the figure is the actual error coverage after SA. Notice that the minimal error coverage is 0.27 since there are some idle time and OS overhead is not counted towards the error coverage. As the error coverage requirement increases from 0.3 to 0.5, the control cost rises accordingly. This study shows that our approach can enable quantitative tradeoff analysis between error coverage and control cost for designers. 

\section{Conclusion}
\label{sec:con}

In this work, we present a novel approach for improving system fault tolerance by leveraging weakly-hard constraints. Our approach includes novel control analysis and scheduling analysis methods under deadline misses, and a meta-heuristic for exploring the design space. Experimental results demonstrate its effectiveness in improving fault tolerance and enabling system-level tradeoffs between control cost and error coverage.

\newpage

\bibliographystyle{ACM-Reference-Format}
\bibliography{reference}

\end{document}